\documentclass[%
reprint,
superscriptaddress,
 amsmath,amssymb,
 aip,
jcp,
]{revtex4-1}

\usepackage{graphicx}
\usepackage{dcolumn}
\usepackage{bm}

\newcommand{\ket}[1]{\vert{#1}\rangle}
\begin{document}


\title{Constrained Adiabatic Trajectory Method (CATM): a global integrator for explicitly time-dependent Hamiltonians}
%
\author{A. Leclerc}
\email{Arnaud.Leclerc@utinam.cnrs.fr}
\affiliation{Institut UTINAM, CNRS UMR 6213, Universit\'e de Franche-Comt\'e, Observatoire de Besan\c con,\\
41 bis Avenue de l'Observatoire, BP 1615, 25010 Besan\c con cedex, France}
\author{G. Jolicard}
\affiliation{Institut UTINAM, CNRS UMR 6213, Universit\'e de Franche-Comt\'e, Observatoire de Besan\c con,\\
41 bis Avenue de l'Observatoire, BP 1615, 25010 Besan\c con cedex, France}
\author{D. Viennot}
\affiliation{Institut UTINAM, CNRS UMR 6213, Universit\'e de Franche-Comt\'e, Observatoire de Besan\c con,\\
41 bis Avenue de l'Observatoire, BP 1615, 25010 Besan\c con cedex, France}
\author{J. P. Killingbeck}
\affiliation{Centre for Mathematics, University of Hull, Hull HU6 7RX, UK}
\affiliation{Institut UTINAM, CNRS UMR 6213, Universit\'e de Franche-Comt\'e, Observatoire de Besan\c con,\\
41 bis Avenue de l'Observatoire, BP 1615, 25010 Besan\c con cedex, France}


\begin{abstract}
The Constrained Adiabatic Trajectory Method (CATM) is reexamined as an integrator for the Schr\"odinger equation. An initial discussion places the CATM in the context of the different integrators used in the literature for time-independent or explicitly time-dependent Hamiltonians. The emphasis is put on adiabatic processes and within this adiabatic framework the interdependence between the CATM, the wave operator, the Floquet and the $(t,t')$ theories is presented in detail. Two points are then more particularly analysed and illustrated by a numerical calculation describing the $H_2^+$ ion submitted to a laser pulse. The first point is the ability of the CATM to dilate the Hamiltonian spectrum and thus to make the perturbative treatment of the equations defining the wave function possible, possibly by using a Krylov subspace approach as a complement. The second point is the ability of the CATM to handle extremely complex time-dependencies, such as those which appear when interaction representations are used to integrate the system.
\end{abstract}
\maketitle

\section{Introduction}

The numerical solution of the Schr\"odinger equation $i\hbar \partial \Psi/\partial t = H \Psi$ is a central element in the understanding of experiments which involve molecular collisions or interactions between molecules and electromagnetic fields. For energy-resolved experiments, stationary theories are used. 
Thus, in the case of quantum diffusion theory, the close-coupling formalism\cite{lester} and the Lippmann-Schwinger approach\cite{lippmann} solutions to the time-independent Schr\"odinger equation $H\Psi = E \Psi$ are sought which correspond to a precise total energy $E$ and also to precise asymptotic conditions and which are found
by integrating differential or integral equations.

In the case of time-resolved experiments, or when the information on the asymptotic solutions provided by the close-coupling techniques is not sufficient (such as in laser control problems), time-dependent treatments are favoured. When the dynamics is driven by a time-independent Hamiltonian, several algorithms can be used to propagate the wave packet which represents the system in Hilbert space.
These include the algorithms presented by Leforestier et al\cite{leforestier}, the second-order differencing scheme (SOD), the split operator method and the short iterative Lanczos propagation. Most of these methods can also be used when the Hamiltonian explicitly depends on time.
However in these cases, the length of the integration steps must be reduced in order to handle any rapid time variations of the hamiltonian matrix.
The propagation scheme is then based on the decomposition of the evolution operator into small increments of duration $\Delta t$:
\begin{equation}
U(t,0)=\prod_{n=0}^{N-1}U((n+1)\Delta t,n\Delta t)
\label{eq1}
\end{equation}
where $\Delta t =t/N$ and
\begin{equation}
U(t+\Delta t, t)=\exp [-(i/\hbar)H(t+\Delta t /2)\Delta t].
\label{eq2}
\end{equation}
Propagation errors due to this scheme are proportional to $(\Delta t)^3$ and involve commutators of the Hamiltonian at successive times. These errors cancel out when the $H$ matrix does not depend on time but there are also errors due to the approximate calculation of the action of 
$\exp [-(i/\hbar)H(t+\Delta t /2)\Delta t]$ on the wave function $\Psi (t)$. 
Thus in the three-point SOD scheme which is based on the equation
\begin{equation}
\Psi(t+\Delta t) \approx \Psi(t-\Delta t) -2i \Delta t H\Psi(t)/\hbar
\label{sod}
\end{equation}
the propagator is conditionally stable and the accumulated error per time step is equal to
\begin{equation}
\mathbf{error} \approx \frac{(\delta t E_m)^3}{3\hbar^3}
\end {equation}
where $E_m$ is the eigenvalue of the discretized Hamiltonian with the largest modulus.

In field-matter coupling problems, the difficulty arising from the presence of high frequencies can be circumvented by introducing the Rotating Wave Approximation (RWA)\cite{shore}. However this approximation generates rather high error plateaus when the step $\Delta t$ becomes too small, which is the case for intense laser fields\cite{kormann}.
Difficult combination of slow quasi-adiabatic evolutions on long time scales together with rapid partial or localised time variations induce very large spectra for the hamiltonian matrix
(e.g. in the theory of a radiative association experiment involving cold molecules fragments).
Such cases create difficult problems of error accumulation for all integrators, although the amplitude and the distribution of the resulting errors in the spectrum is not the same for them all.

To obtain high-accuracy integrators for multi-dimensional systems evolving adiabatically, one can also introduce the symplectic partitioned Runge-Kutta methods. 
Using the work of Gray and Verosky on real and time-independent Hamiltonians\cite{gray},
Sanz-Serna et Portillo\cite{sanz} have generalized the method to time-dependent Hamiltonians by transforming the system into an autonomous equation by introducing an additional conjugate pair of variables $(\mathcal{P},\mathcal{T}=t)$. Another accurate time
propagation method for an explicitly time-dependent Hamiltonian have been produced by Kormann\cite{kormann}, by replacing the Hamiltonian $H$ in Eq.\eqref{eq2} by a suitable truncation of the Magnus series\cite{magnus,talezer} $\overline{H}$ and by using the short iterative Lanczos scheme for computing the matrix-vector multiplication $\exp [-(i/\hbar)\overline{H}\Delta t] \Psi$.

While the stability of an integrator, its accuracy and its ability to conserve the norm of the wave function are important features, it is also necessary to consider other elements such as the calculation time needed for a given accuracy, the required memory capacity, the complexity and generality of the integrator and also any constraints which could prevent its use in some cases. For instance some integrators such as the SOD cannot handle non-hermitian Hamiltonians\cite{leforestier}. The split operator scheme\cite{feit} requires that the kinetic operator does not mix coordinates and their associated momenta. The multi-configuration time dependent Hartree (MCTDH) method\cite{mctdh,mctdh2,mctdh3}
requires important preliminary work to rewrite the kinetic and potential operators, while the calculation of higher-order terms of the Magnus development is a complicated task which is only tractable if the couplings have separated time and coordinate dependencies\cite{kormann}.\\

In this article we investigate the performances of the Constrained Adiabatic Trajectory Method (CATM) \cite{CATM,CATM2,CATM3} as a global integrator for the Schr\"odinger equation, with particular emphasis on a system which is adiabatic, in the sense that the system is correctly described by small subspaces spanned by eigenvectors of the molecular Hamiltonian $H(x,t)$, or by Floquet eigenstates of the field-dressed Hamiltonian.
The CATM is well suited to the description of systems driven by Hamiltonians with explicit and complicated time variations. This method does not have cumulative errors and the only error sources are the non-completeness of the finite molecular and temporal basis sets used, and the imperfection of the time-dependent absorbing potential which is essential to impose the correct initial conditions.

In sec. \ref{secII}, the CATM theory is placed in context with regard to other treatments such as the Floquet theory or the $(t,t')$ method, with emphasis on the concept of adiabaticity and on the compatibility between this concept and the time-dependent wave operator theory.
Then three points are particularly studied, all related to the fact that the CATM proposes a global integrator for explicitly time-dependent Hamiltonians and thus is not in the category of methods described by Eq.\eqref{eq2}.
In sec. \ref{comparaison} we present some comparisons between the CATM, the SOD scheme and the split-operator method.
The second question we ask in sec. \ref{krylov} is how the influence of a Krylov growing subspace algorithm \cite{krylovcambridge} directs the convergence properties of the CATM as compared to a perturbative recursive distorted wave approximation (RDWA) approach to solve the wave operator equations\cite{reviewgeorges2} (with or without absorbing potential, because the expansion of the Floquet spectrum under the influence of the absorbing potential\cite{CATM2} also directly acts on the convergence properties).
In sec. \ref{reint}, another important point emerges from a numerical problem that we have noted in previous CATM calculations \cite{CATM3}, in the case of a multistep propagation (if the time interval is too long to be treated with only one global step, it can be divided into several large steps treated in succession). Using the absorbing operator sometimes leads to high-frequency parasites characteristic of the Gibbs phenomenon. We then try to evaluate the benefits of introducing an interaction representation before applying the CATM to implement the time propagation. This provides a new test for the method, in the presence of general time variations in the Hamiltonian. 
Sec. \ref{conclusion} is devoted to the conclusion.



\section{Stationary and dynamic treatments for adiabatic processes in molecular physics \label{secII}}


The main difficulty in studying adiabatic processes comes from the fact that in most cases adiabatic or even quasi-stationary molecular interactions are combined with fast time variations. A purely stationary process where the time only appears as a global phase in the wave function implies that $H$ is self-adjoint and time-independent.  On the other hand, as soon as the Hamiltonian is non-self-adjoint and acquires resonance states, characteristic times appear, such as the lifetime of the initial state or a characteristic tunneling passage time\cite{NHQM,time}.

Things are even more complicated when the Hamiltonian becomes explicitly time-dependent, either because certain classical degrees of freedom are present and are coupled with quantum degrees of freedom or because interaction representations are used during the calculation.
There is now no simple expression for the evolution operator and the Dyson expansion in powers of the Hamiltonian is no longer consistent with Eq.\eqref{eq1}, except if we use a Magnus expansion with a time step $\Delta t$ which depends on the order of the Magnus expansion.

The difficulty of constructing the Magnus series can be circumvented by using the $(t,t')$ theory\cite{peskin}. Thus the Schr\"odinger equation for a time-dependent Hamiltonian can be solved in the same way as that for a time-independent Hamiltonian by working within the extended Hilbert space ${\cal K}$. This extended space was first introduced by Sambe\cite{sambe} for periodic Hamiltonians and was then generalized by Howland\cite{howland}. In short, the $(t,t')$ method solves the Schr\"odinger equation
\begin{equation}
i\hbar \frac{\partial}{\partial t} \Psi(x,t)=H(x,t)\Psi(x,t)
\label{schrodinger}
\end{equation}
by adding a new variable $t'$ to define the extended Hilbert space. The corresponding wave function
$\overline{\Psi}(x,t',t)$ is related to $\Psi(x,t)$ by
\begin{equation}
\Psi(x,t)=\overline{\Psi}(x,t',t)|_{t'=t},
\end{equation}
with
\begin{equation}
\overline{\Psi} (x,t',t)=\exp[-(i/\hbar)H_F(x,t')(t-t_o)]\overline{\Psi}(x,t',t_o)
\end{equation}
where $H_F$ is a Floquet-type operator:
\begin{equation}
H_F(x,t')=H(x,t')-i\hbar \frac{\partial}{\partial t'}.
\end{equation}
The choice between two possible representations of the initial state depends on the initial process
studied. Depending on the circumstances one can choose a time-independent initial state,
\begin{equation}
\overline{\Psi}(x,t',t_o)=\Psi(x),
\end{equation}
or an initial state which is well-defined for a specific initial time $t_o$,
\begin{equation}
\overline{\Psi}(x,t',t_o)=\delta(t'-t_o)\Psi(x).
\end{equation}
We thus find an integration scheme which is based on Eqs.\eqref{eq1} and \eqref{eq2} but with a Hamiltonian which belongs to the larger extended Hilbert space. This can possibly create memory capacity problems, especially when the $t'$ interval is very large.

In the periodic case, i.e. $H(t)=H(t+T)$ with $T=2\pi/\omega$, one can transform the dynamic problem into an equivalent time-independent infinite-dimension eigenvalue problem\cite{shirley} and generalize it to the complex quasivibrational energy formalism by including finite ${\cal L}^2$ representations
of the molecular continua\cite{chu}. Using the quantum variable $\theta=\omega t$ and introducing the Floquet Hamiltonian:
\begin {equation}
H_F(\theta)=H(\theta) -i\hbar\omega \frac{\partial}{\partial \theta}
\end{equation}
which is defined in the enlarged Hilbert space ${\cal K}={\cal H} \otimes {\cal L}_2 (d\theta/2\pi)$,
the evolution operator acting in the enlarged space then becomes
\begin{equation}
U_{H_F}(t,t_o)=\exp [-iH_F(t-t_o)/\hbar]
\label{eq12}
\end{equation}
and is related to the evolution operator in the Hilbert space $U_H(t,t_o)$ as follows:
\begin{equation}
U_{H_F}(t,t_o)=\tau_{-\omega t}U_H(t,t_o)\tau_{\omega t}
\end{equation}
where $\tau_{\omega t}=e^{i\omega t \partial/\partial \theta}$ is a phase translation operator which acts on the functions of ${\cal L}_2$. We thus obtain a direct relation between the standard solution $\Psi$ in the Hilbert space ${\cal H}$ and the solution $\overline{\Psi}$ in the extended space~${\cal K}$:
\begin{equation}
\Psi(x,t)=\tau_{\omega t}\overline{\Psi}(x,t,\theta)=\overline{\Psi}(x,t,\theta+\omega t).
\end{equation}
This formulation also establishes a connection between the quantum and the semi-classical formalisms for field-matter interactions at the intense field limit\cite{reviewguerin}.
It also provides a way to describe the mix of adiabatic and sudden effects in some experiments by using several time scales.

However, the $(t,t')$ and the Floquet theories are handicapped by having quite large memory requirements in numerical applications. Moreover, in the enlarged Hilbert space, the calculation of the action of the operator $\exp [-iH_F(t-t_o)/\hbar]$ [Eq.\eqref{eq12}] on the initial state remains a delicate problem when the $H_F$ spectrum is very dispersed, even if the Chebyshev global scheme can be used.

However, another approach is possible. At the adiabatic limit, we can stop searching for exact numerical solutions by adopting an adiabatic approximation such as\cite{berry}:
\begin{equation}
\Psi (t) \approx \exp \{\frac{1}{i\hbar}\int^t_{t_o}E(t')dt'-\int^t_{t_o}\langle \phi(t')|\partial\phi(t')/\partial t'\rangle dt'\} \phi(t)
\label{eq15}
\end{equation}
where $\phi (t)$ is an instantaneous eigenvector:
\begin{equation}
H(t)\phi(t)=E(t)\phi(t),
\label{eigenvalue}
\end{equation}
and where the initial wave function $\Psi (t_o)$ is assumed to be equal to the instantaneous eigenvector $\phi (t_o)$. The main weak point of this approach is that it is rigorous only at the purely adiabatic limit. Such a case is exceptional and in most cases, the dynamics generates non-adiabatic couplings which mix several eigenvectors. Adiabatic formulae such as Eq. \eqref{eq15} must then be generalized by introducing degenerate active spaces and non-abelian geometric phases. 
In the last part of this section we demonstrate that the wave-operator is undoubtedly the better framework to describe this generalization. Indeed the non-adiabatic couplings are generated by the operator $i\hbar \partial / \partial t$ and if one renders zero this operator in the fundamental equation which defines the time-dependent wave operator one obtains the fundamental equation which defines the stationary wave operator. The stationary form is then the pure adiabatic limit of the time-dependent form. In the following we will denote by $\Longrightarrow$ the passage from the time-dependent to the stationary equations induced by this pure adiabatic limit and by $\not\Longrightarrow$ the non-passage. Evidently one cannot go from the Schr\"odinger equation \eqref{schrodinger} to the eigenvalue equation \eqref{eigenvalue} by setting $\partial/\partial t \rightarrow 0$ in the first one,
\begin{equation}
(H(x,t)-i\hbar\frac{\partial}{\partial t}) \Psi (x,t)=0 \not\Longrightarrow (H(x,t)-E(t))\phi(x,t)=0,
\label{eq17}
\end{equation}
because the operator $i\hbar \partial/\partial t$ also generates the dynamic phase which is associated with $E$ and which is an integral part of the wave function. 

These drawbacks disappear within the framework of the wave operator theory because the time-dependent wave operators does not include the dynamic phase, which is factorized separately. We define $P_o$ as the projector corresponding to the finite group of non-perturbed eigenvectors which mix under the influence of non-adiabatic couplings, $Q_o$ being the projector onto the complementary space. Let $P_t$ be the projector associated with the corresponding group of perturbed eigenvectors at the time $t$ when the Hamiltonian takes the value $H(t)$ and $U(t,0,H)$ be the evolution operator. The Bloch wave operator is defined by\cite{bloch,reviewgeorges1}
\begin{equation}
\Omega^t=P_t(P_oP_tP_o)^{-1}=P_o+Q_oX^tP_o
\label{blochwaveoperator}
\end{equation}
(The inverses are defined within the subspace $S_o$, $(P_oP_tP_o)^{-1}$ is the inverse of $P_t$ within the space $S_o$),
and the time-dependent wave operator is defined by\cite{reviewgeorges2}
\begin{eqnarray}
\Omega(t,0)&=&U(t,0;H)(P_oU(t,0;H)P_o)^{-1} \nonumber \\
&=&P_o+Q_oX(t,0)P_o.
\label{tdwaveoperator}
\end{eqnarray}
These operators define a generalized adiabatic framework in which the wave function is an instantaneous linear combination of the eigenvectors spanning the subspace with projector $P_t$. The time-dependent wave operator factorizes the dynamic phase and the non-Abelian Berry phase inside $U(t,0;H_{\text{eff}})$\cite{viennot2006}:
\begin{eqnarray}
U(t,0;H)P_o&=&\Omega(t,0)U(t,0;H_{\text{eff}}) \\
\mathrm{with} \; H_{\text{eff}}(t)&=&P_oH(t)\Omega(t,0)P_o
\end{eqnarray}
These phases (which express rapid evolutions generated by the non-adiabatic couplings) are thus separated from the adiabatic evolution, the latter being included in $\Omega$. 
The result then is that the fundamental equations for $\Omega(t,0)$ [Eq.\eqref{tdwaveoperator}] and $\Omega^t$ [Eq.\eqref{blochwaveoperator}] adiabatically correspond to each other by cancelling out the operator $\partial/\partial t$. Instead of the non-implication \eqref{eq17} we now have:
\begin{widetext}
\begin{equation}
Q_o(1-X(t,0))H(t)(1+X(t,0))P_o =i\hbar \partial X(t,0)/\partial t \, \Longrightarrow \, Q_o(1-X^t)H(t)(1+X^t)P_o =0
\label{eq20}
\end{equation}
\end{widetext}
A direct consequence is that the adiabatic limit of the time-dependent wave operator is
given by a succession of instantaneous Bloch wave operators\cite{viennot2005}.

The wave operator theory is compatible with both the $(t,t')$ method and the Floquet theory, and the remarkable property of Eq.\eqref{eq20} at the adiabatic limit is conserved if we work in the enlarged Hilbert space ${\cal K}$. Thus, if $H(t)$ is $T=2\pi/\omega$ periodic ($T$ can be arbitrarily large), the two fundamental equations which define $\Omega(t,0)$ and $\Omega^t$ in the ${\cal K}$ space can be derived (with $t$ as a quantum variable and no longer a parameter). Denoting the wave operator within the ${\cal K}$ space by $\Omega$ we have the implication
\begin{equation}
\begin{array}{c}\Omega(H(t)-i\hbar\partial/\partial t)\Omega \\
=(H(t)-i\hbar \partial /\partial t)\Omega 
\end{array} \Longrightarrow
\begin{array}{c} \Omega^t H(t) \Omega^t \\
=H(t)\Omega^t.
\end{array}
\label{eq21}
\end{equation}
We note that the equations become identical by transforming $H_F(t)=H(t)-i\hbar\partial/\partial t$ into $H(t)$, i.e. by neglecting the time-derivative operator 
(however on the left hand side $t$ is a true coordinate of the ${\cal K}$ space, while it is only a fixed parameter on the right hand side.)

Nevertheless there is not a perfect equivalence between the two left-hand sides of Eqs.\eqref{eq20} and \eqref{eq21}.
The equation: $i\hbar\partial X(t,0)/\partial t=Q_o(1-X(t,0))H(t)(1+X(t,0))P_o$ is a non-linear evolution equation within the Hilbert space, which uses an imposed initial $X(t=0,0)$ which is consistent with the chosen initial wave function. Integration of the equation indirectly gives the wave function $\Psi(x,t)$, since the initial conditions are automatically satisfied.
By contrast the left part of Eq.\eqref{eq21}, $\Omega  H_F(t) \Omega=H_F(t) \Omega $, is a stationary eigenvalue equation within the extended Hilbert space and its solution gives Floquet eigenvectors with an intermediate normalization. For a degenerate $S_o$ space, no column of $\Omega$ has {\it{a priori}} an initial value compatible with the initial wave function. For intense couplings, a large number of columns will unfortunately be necessary to approach the solution of the Schr\"odinger equation (so that a subspace $S_o$ with projector $P_o$ of high dimension will be necessary). The left equation in \eqref{eq21} then provides $\Omega$ in a global way, but contrary to its counterpart in Eq.\eqref{eq20}, which is amenable to a step-by-step integration, it is inapplicable to integrate the time-dependent Schr\"odinger equation.

To overcome this major difficulty, within the framework of the CATM\cite{CATM} a complex absorbing potential is added to the Hamiltonian $H(t)$ during an extension of the integration interval, in such a way as to impose an initial vector $\langle t=0|\Omega$ compatible with the initial value of the wave function $\Psi (t=0)$. This can be done using only a non-degenerate subspace $S_o, (P_o)$. 
Within a one dimensional subspace the wave operator representation becomes a column.
In a previous paper\cite{CATM3}, we have studied the matrix expressions of a general time-dependent absorbing potential ${\cal V}$ for any initial wave function $\Psi (x,t=0)$. In this framework, the integration of
\begin{equation}
\Omega  (H_F(t) +{\cal V}(t)) \Omega = (H_F(t) +{\cal V}(t))\Omega 
\end{equation}
gives, in a complete basis for the ${\cal K}$ space, the expression of the column $|\Omega \rangle$ which is a Floquet eigenvector $|\lambda \rangle$ belonging to the first Brillouin zone (with an intermediate normalization), i.e.
\begin{equation}
|\Omega \rangle= |\lambda \rangle /\langle i,n=0|\lambda \rangle
\end{equation}
where $|i,n=0\rangle$ is the basis vector for ${\cal K}$ in the first Brillouin zone and is chosen to have maximum overlap with the initial wavefunction. $|\lambda\rangle$ is a Floquet eigenvector such that
\begin{equation}
(H_F+{\cal V}) |\lambda\rangle=E_{\lambda}|\lambda\rangle
\end{equation}
where ${\cal V}$ ensures the equality between $\langle t=0|\lambda\rangle$ and the initial wave function.
Eventually the wave function is simply given by
\begin{equation}
|\Psi (t)\rangle=\exp^{-iE_{\lambda}t/\hbar} \langle t|\lambda \rangle .
\label{wavefunction}
\end{equation}

This formulation then gives a global solution for the Schr\"odinger equation with an explicitly time-dependent Hamiltonian. This generalizes finite-basis treatments within the Hilbert space ${\cal H}$ to finite-basis treatments within the extended Hilbert space ${\cal K}$. The method uses, in addition to the radial complex absorbing potential\cite{CAP1,CAP2} (which is necessary to reveal resonances), extra absorbing operators asymptotically placed along the time axis to impose the initial conditions. Contrary to that of Eq.\eqref{eq15} where the non-adiabatic couplings are neglected, this solution is rigorous even if it is based on an adiabatic hypothesis. The addition of a time-dependent absorbing operator makes the wave function proportional to a single Floquet eigenvector [Eq.\eqref{wavefunction}] which is calculated in an iterative way within the framework of the wave operator theory.

Similar concepts have been introduced long ago by Peskin et al \cite{peskin2,peskin3} in the framework of the $(t,t')$ theory, for the calculation of Green functions within the extended space ${\cal K}$. Introducing a $t'$-dependent absorbing potential to impose the boundary condition for the $t'$ axis, they have shown that it was possible to replace the time-dependent Schr\"odinger equation by an inhomogeneous time-independent linear system and thus to calculate transition probabilities using a scattering matrix formalism. They solved the linear system using a Krylov subspace-based iterative method in combination with a Fourier grid preconditioner. Despite the numerous similarities between this approach and ours, the fundamental working equations and the iterative procedure we use are different.


From a more pragmatic point of view, the aim of the following sections 
is to give some comparisons between the CATM and other schemes and to analyse in detail how the presence of the absorbing operator affects the several possible variants in the method used to determine the eigenvector $\ket{\lambda}$.

\section{Comparisons between the CATM, the SOD scheme and the split-operator method \label{comparaison}}

In Sections \ref{comparaison}, \ref{krylov} and \ref{reint}, we make numerous test simulations with the well-known example of the photodissociation of the molecular ion $H_2^+$ within the Born-Oppenheimer approximation. 
This example is a simple 1D dynamical system but it constitutes a significant test because the CATM is not focused, as MCTDH, on the treatment of multi-dimensional quantum systems with the use of efficient time-dependent basis-set. The CATM proposes a new scheme to integrate the dynamics driven by a complicated and fast time-evolution. These two aspects (large dimensions and complicated time-evolution) are nevertheless sometimes correlated. The use of intermediate representations allows us to reduce the basis set dimensions, but in so doing it makes the time-dependence complicated (cf. Section \ref{reint}). In the present section, a strong adiabatic laser-field envelope with several hundred optical oscillations is studied in order to test the capacity of our model to reproduce such extreme adiabatic situations.

\subsection{Model for $H_2^+$ illuminated by intense pulses}

\begin{figure}[!ht]
 \centering
\includegraphics[width=\linewidth]{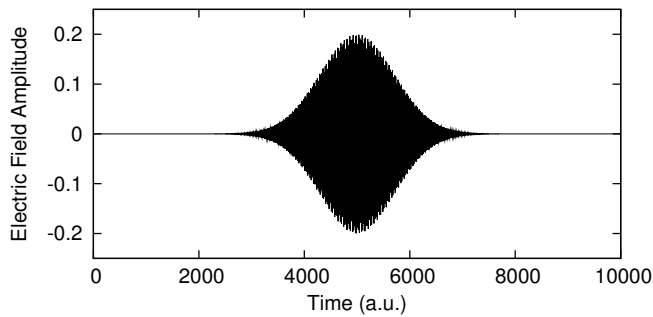}
 \caption{Adiabatic laser pulse with angular frequency $\omega=0.2958678$~au and total duration 10000 au (0.24ps) with a gaussian shape.}
\label{pulseadiabatic}
\end{figure}

\begin{figure}[!ht]
 \centering
\includegraphics[width=\linewidth]{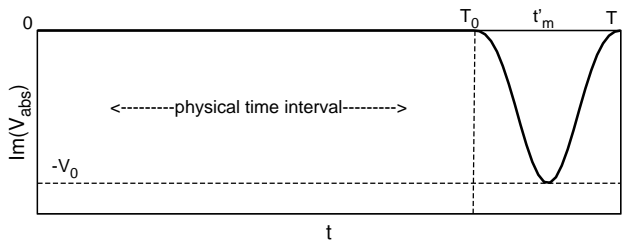}
 \caption{Imaginary part of the time-dependent complex absorbing potential defined in Eq. \eqref{functionvabs}.}
\label{absshape}
\end{figure}

We study the nuclear vibration and we only take into account the first two effective potentials\cite{bunkin} in the two lowest electronic states $^2\Sigma_g^+$ and $^2\Sigma_u^+$.
The nuclear Hamiltonian is the sum of two terms
\begin{equation}
 H=H_0+W(t).
\end{equation}
Here $H_0=K+V_0(x)$ is the field-free Hamiltonian of $H_2^+$, which is pre-diagonalized on a radial grid basis using a grid  method \cite{grid} in the presence of a radial complex absorbing potential \cite{CAP1,CAP2}, to obtain the vibrational eigenbasis with 200 eigenvalues $\{E_j\}$ and bi-orthogonal eigenstates $\{\ket{j}\}$, as well as the electric moment operator represented by the matrix $\mu_{ij}=\langle i \vert \mu \ket{j}$. We then calculate the dynamics in the presence of a semi-classical intense time-dependent electric field.
In our example the electric field envelope is given by the gaussian function
\begin{equation}
 E(t)=E_0 \exp\left( -\left(\frac{t-t_m}{\tau}\right)^2 \right)
\label{gaussienne}
\end{equation}
with $\tau=1000$ au (cf. Fig. \ref{pulseadiabatic}).
The total duration of the adiabatic pulse is $T_0=2t_m=10000$ au (i.e. 0.24 ps) and the carrier wave angular frequency $\omega$ is 0.2958678~au, which corresponds to a wavelength of 154 nm.
Within the framework of the dipole approximation, the coupling term $W(t)$ is 
\begin{equation}
 W(x,t)=\mu(x) . E(t).
\end{equation}
The CATM is then based on the Floquet eigenproblem
\begin{equation}
 \left[ H_0(x)+W(x,t)+{\cal V}(x,t) - i\hbar \frac{\partial}{\partial t}\right] \ket{\lambda} = E_{\lambda} \ket{\lambda},
\end{equation}
${\cal V}$ being a time-dependent absorbing operator present only on the supplementary interval $t\in[T_0,T]$ and defined \cite{CATM3} in Table~\ref{derniere}.
The absorbing interval is $\Delta T\simeq3600$~au with a centred bell shape for the absorbing operator (cf. Fig.\ref{absshape}),
\begin{equation}
 V_{abs}(t)=-i \; V_0 \; \text{sinc} ^2 \left( \frac{(t - t_m')}{\Delta T} \right),
\label{functionvabs}
\end{equation}
with $t_m'=T_0+\frac{\Delta T}{2}$.

\begin{table}[htp]
\centering
\begin{center}
\renewcommand{\arraystretch}{2}
\begin{tabular}{ccccc}
\hline
\hline $V_{abs}(t_i)$ & 0 & 0 & (column $l$) & 0 \\
0 & $V_{abs}(t_i)$ & 0 & $-\frac{\langle j \vert \Psi (0) \rangle}{\langle l \vert \Psi(0) \rangle} \times 
\left( V_{abs}(t_i) + E_j - E_l \right)$ & 0 \\ 
0 & 0 & $\ddots$ & $\vdots$ & 0 \\ 
0 & 0 & 0 & 0 (row $l$) & 0 \\ 
0 & 0 & 0 & $-\frac{\langle j \vert \Psi (0) \rangle}{\langle l \vert \Psi(0) \rangle}\times 
\left( V_{abs}(t_i) + E_j - E_l \right)$ & $V_{abs}(t_i)$ \\
\hline
\hline
\end{tabular}
\end{center}
\caption{Matrix representation of one block $t_i$ of the absorbing potential within the bi-orthogonal eigenbasis set $\{\ket{j}\}$ of $H_0$.}
\label{derniere}
\end{table}

The fundamental Floquet period is the total duration $T$. The time description can be made using time-periodic functions $\langle t \ket{n}= 1/\sqrt{T} \, e^{- 2 \pi i n t/T}$ ($n\in {\mathbb N},\; n=-N/2 \dots (N/2-1)$) as a finite basis representation (FBR), and the associated discrete variable representation (DVR) is defined by $\ket{t_i}=1/N \sum_{n=-N/2}^{N/2-1} e^{- 2 \pi i n (t-t_i)/T}$. $N$ is the number of Fourier basis functions, or equivalently the number of grid points which describe the time dimension. The required extended Hilbert space can be of quite large dimension if one or other component (${\cal H}$ or ${\cal L}^2$) has a large dimension. The calculation of $\ket{\lambda}$ can be efficiently undertaken using the wave operator theory in the case of a one-dimensional active space. We then have to solve Eq.\eqref{waveoperator} (see Appendix), $\Omega$ being simply proportional to the eigenvector.
Eventually the transition probabilities $P(\ket{j},t)=|\langle j \ket{\Psi(t)}|^2$ as well as the dissociation probability, 
\begin{equation}
 P_{\text{diss}}=1-\sum_{\text{bound states}}|\langle j \ket{\Psi(t)}|^2,
\end{equation}
can be calculated.

In terms of memory, the CATM mainly requires the storage of a vector of the extended Hilbert space, which is modified iteratively. Because of the use of FFT, the computational effort of the program with respect to the time propagation scales as $2N_{iter} N \log (N)$ where $N$ is the number of Fourier basis functions and $N_{iter}$ the number of iterations needed to converge in the wave operator calculation (see Appendix \ref{app1}). The minimum $N$ is roughly proportional to the total duration for a given maximum frequency of the field. If $N$ is sufficient to stabilize the fast Fourier transforms, then the accuracy of the results does not depend directly on $N$ but rather on the absorbing potential parameters. $N_{iter}$ (10-40 in general) does not depend on the dimension of the matrix $H_F$. 

\subsection{Comparative integration schemes}

The CATM is a global propagator for explicitly time-dependent Hamiltonians. However, since the Chebyshev polynomial development of the evolution operator $\exp(H_F(t-t_0)/(i\hbar))$ in the extended space offers another very precise global solution\cite{kosloffchebyshev}, it appeared worthwhile first to compare these two global solutions. In the Chebyshev approach the global integrator is constructed on an iterative way and requires at least a minimum number $N_{Cheb}$ of iterations given by:
\begin{equation}
N_{Cheb}>\frac{\Delta E \;\; T}{2 \hbar},
\end{equation}
where $\Delta E$ is the complete energy range (here the Floquet energy range) and $T$ the total duration. In the present case, we have
\begin{equation}
\Delta E \simeq \text{Max}(E_i) + N \frac{2 \pi}{T}.
\end{equation}
With $\text{Max}(E_i)\simeq 2.934$a.u. (i.e. the maximum value of the $H_0$ spectrum) and $N=2048$, we obtain a minimum number of iterations $N_{Cheb} = 21084$. Exactly as for the CATM, each iteration of the Chebyshev polynomial scheme requires the multiplication of $H_F$ by a vector of the extended space. By comparison, the CATM in combination with the wave operator theory generally converges within only a few tens of iterations under the same conditions. The CATM appears then as largely competitive compared to the Chebyshev scheme.

It is also interesting to compare the CATM with two non-global propagators, namely the SOD scheme given in Eq.\eqref{sod} and the split-operator method. To construct a significant comparison, the representation of the molecular $H(t)$ for both the CATM and the SOD is made on the same molecular basis, the eigenbasis of $H_0$. Thus we can use the simplest form of absorbing potential for the CATM if the initial state is an eigenstate of $H_0$. Of course, by doing this, we lose the advantage of the more sparse hamiltonian matrix representation when it is represented using a DVR grid basis for $x$. With the SOD, the accumulated error per time step is proportionnal to $\delta t^3$ where $\delta t=T/N_{sod}$ is the time step. Thus after $N_{sod}$ propagation steps if we wish to have a final fixed error smaller than a given number $e$, i.e.
\begin{equation}
N_{sod} \times \left( \alpha \frac{T}{N_{sod}} \right) ^3 < e,
\end{equation}
we must choose $N_{sod}$ greater than $\left( \alpha^{3/2} \frac{T^{3/2}}{\sqrt{e}} \right)$.

The problem can also be studied with the split-operator method, which requires $H(t)$ to be represented on a DVR grid basis for $x$. This scheme is based on a splitting of the kinetic and potential energies in the evolution operator for each step\cite{feit}:
\begin{eqnarray}
\exp\left(-\frac{i}{\hbar}H \delta t\right)&=&\exp\left(-\frac{i}{\hbar}(K+V) \delta t\right) \nonumber \\
&\simeq& e^{- i K \delta t / (2 \hbar) } e^{-i V \delta t / \hbar } e^{- i K \delta t / (2 \hbar) },
\end{eqnarray}
where $V=V_0(x)+W(x,t)$. 
In the present case, a second splitting is necessary for the potential term because $W(x,t)$, which represents the coupling terms between the two electronic effective potential curves, does not commute with $V_0$:
\begin{equation}
\exp\left(-\frac{i}{\hbar}V \delta t\right) \simeq e^{- i V_0 \delta t / (2 \hbar) } e^{-i W \delta t / \hbar } e^{- i V_0 \delta t / (2 \hbar) }.
\end{equation}
In the 1-D $H_2^+$ model, the kinetic energy is diagonal in the momentum representation (FBR), and the potential energy is block-diagonal in the coordinate representation (DVR grid for $x$). More precisely, $\exp(-iV_0\delta t /\hbar)$ is diagonal in the two central blocks corresponding to the two surfaces. $\exp(-iW\delta t/\hbar)$ possesses a diagonal representation equal to $\cos(W\delta t /\hbar)$ in the same central blocks and a diagonal representation equal to $i\sin(W \delta t /\hbar)$ in the two off-diagonal blocks. Of course this property will reduce the split-operator computational effort with respect to the other two techniques, which are implemented on the eigenbasis of $H_0=T+V_0(x)$ where the coupling $W(x,t)$ is not block-diagonal (which is in a sense a more general exercise for the methods).

\subsection{Results}

\begin{figure}[!ht]
 \centering
\includegraphics[width=\linewidth]{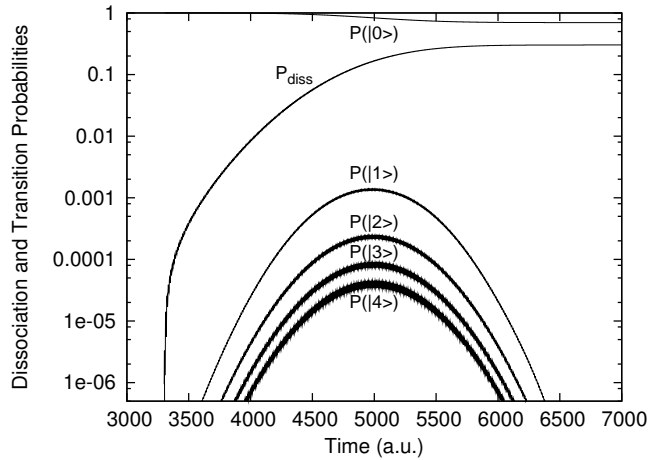}
 \caption{Dissociation and transition probabilities for $H_2^+$ submitted to the pulse of Fig. \ref{pulseadiabatic}. The initial state was the fundamental state $v=0$.}
\label{pdiss1}
\end{figure}

We focus on the transition probabilities to the first bound states and on the dissociation probability. The system is quasi-adiabatic and mainly follows the initial state, as is shown in Fig. \ref{pdiss1}. There are small non-adiabatic transitions during the pulse (for instance $P_{0\rightarrow1} \lesssim 10^{-3}$). Final inelastic probabilities are all smaller than $10^{-15}$.

\begin{figure}[!ht]
 \centering
\includegraphics[width=\linewidth]{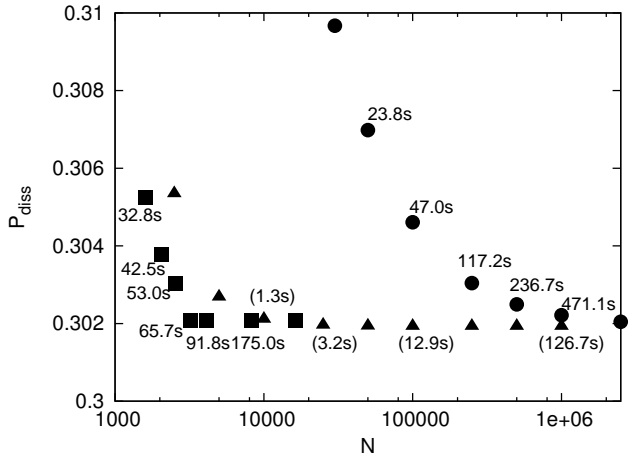}
 \caption{Final dissociation probability given by the CATM (squares), the SOD scheme (rounds) and the split-operator method (triangles) vs number of Fourier basis functions (CATM) or number of time steps (SOD and Split-operator). The associated CPU-time are mentioned near each point. The CATM and the SOD scheme work within the $H_0$ eigenbasis and the split-operator works within a DVR $x$-basis.}
\label{pdiss2}
\end{figure}

The three methods are successively applied to this example with a variable number of time steps (SOD and split-operator) or with a variable number of Fourier basis functions (CATM). Fig.\ref{pdiss2} shows the values of the final dissociation probability given by the CATM and the SOD scheme, both working in the $H_0$ eigenbasis and by the split-operator method working in the DVR grid basis on $x$. We have mentionned CPU-times using a small workstation. For an equivalent accuracy on the final $P_{diss}$, the required time points numbers and CPU times are very different. The CATM is more efficient than the SOD scheme and the split-operator method is the fastest to converge to a correct value of the final dissociation probability.

\begin{figure}[!ht]
 \centering
\includegraphics[width=\linewidth]{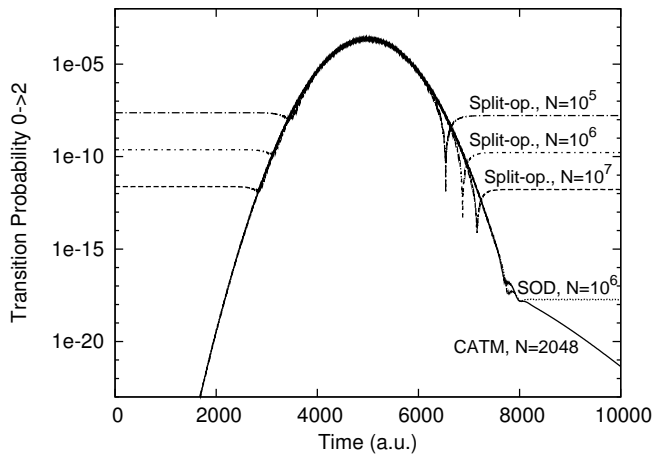}
 \caption{Time evolution of the transition probability $| \langle 2 | \Psi (t) \rangle |^2$ given by the CATM with $N=2048$ Fourier functions (cpu time: 42.5s), by the SOD method with $N=10^6$ time steps (cpu time: 471.1s), by the split-operator scheme with $N=10^5$ time steps (cpu time: 12.9s), with $N=10^6$ time steps (cpu time: 126.7s) or with $N=10^7$ (cpu time: 1244.3s).}
\label{ptrans02}
\end{figure}

Nevertheless in such a quasi-adiabatic problem it is also important to look at the small non-adiabatic transition probabilities. Fig. \ref{ptrans02} shows that the CATM and the SOD scheme give better results for small probabilities than the split-operator (which failed to correctly reproduce the beginning and the end of the dynamics, even if a large number of steps is used). The non-significant decrease at the end of the process given by the CATM comes from an imprecision in the Floquet eigenvalue $E_{\lambda}$ which appears in the exponential of Eq.\eqref{wavefunction}.

\section{The influence of the absorbing potential and of a Krylov subspace procedure on the convergence of the CATM \label{krylov}}

\begin{figure}[!ht]
 \centering
\includegraphics[width=\linewidth]{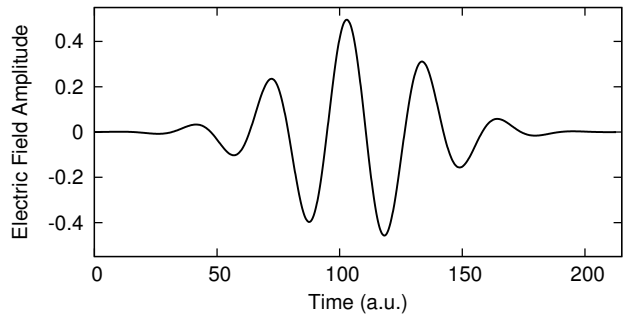}
 \caption{Laser pulse with angular frequency $\omega=0.2958678$~au and total duration 212 au with gaussian shape. All results in the current section relate to this pulse (with variable intensity).}
 \label{champ1}
\end{figure}

From now on we only work with the CATM. In the current section we choose to treat an ultra-short pulse with total duration $T_0=212.9$~a.u. (i.e. 5.15 fs), $\tau=40$ a.u. with the same angular frequency $\omega=0.2958678$~a.u. (cf. Fig. \ref{champ1}). CATM calculations are driven with a small basis of 256 Fourier functions to describe the time evolution. The absorbing interval is $\Delta T=70$~au with a greater amplitude.

For a given duration of the pulse, the different computational features that can be modified are studied successively. These features are :
\begin{itemize}
 \item the choice of the integration technique used to find the wave operator [recursive distorted wave approximation (RDWA) perturbative calculation or RDWA plus Krylov subspace algorithm, see Appendix \ref{app1}];
 \item the amplitude of the time-dependent absorbing potential $V_0$ [cf. Eq.\eqref{functionvabs}]. If $V_0$ is too weak the eigenvector is just a Floquet eigenvector but is not connected to the correct initial condition. If $V_0=0$, this is no longer a CATM calculation.
 \item the maximum amplitude of the electric field $E_0$.
\end{itemize}
Instead of approaching the Floquet eigenvector by adding successive correction terms to a trial vector, the Krylov subspace procedure uses the correction terms to construct a small but growing subspace in which direct diagonalisation gives a better estimate of the Floquet eigenvector. This procedure may improve the convergence properties of the CATM.
In particular, an interesting point is to study how this expected improvement interacts with the expansion of the Floquet spectrum\cite{CATM2} in presence of the absorbing potential; this expansion has an important influence on the convergence properties.

The calculation is halted when the following convergence criterion is satisfied:
\begin{equation}
 \text{norm} \left[ (H_F-E_{\lambda})\ket{\lambda}  \right] < 10^{-12}.
\end{equation}
We count the number of iterations required to reach this level of convergence for the two methods, which indicates the efficiency of the calculation and allows us to estimate the radius of convergence. Here we assume that the initial state is one of the $\{\ket{j}\}$ eigenstates, $\ket{j=i}$. The biggest $(j\neq i)$ component $\epsilon$ at $t=0$ (which should ideally be zero) is also monitored to estimate the quality of the results:
\begin{equation}
 \epsilon=\text{max} \left[\vert \langle j \ket{ \Psi(t=0)} \vert^2 \right] \text{ with } j\neq i.
\label{residue}
\end{equation}

\begin{figure}[!ht]
 \centering
\includegraphics[width=\linewidth]{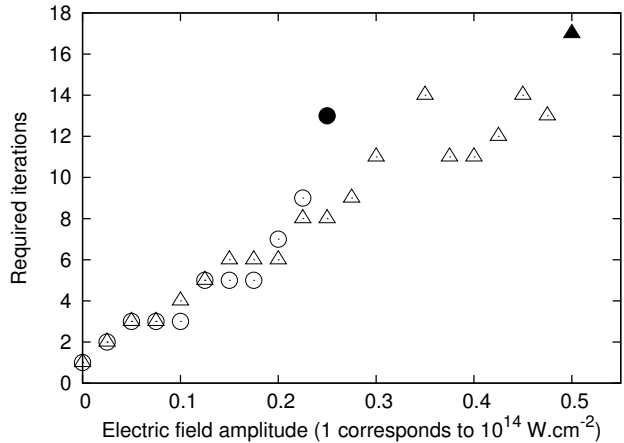}
 \caption{Number of iterations required until convergence vs electric field amplitude $E_0$, with $V_0=0$, using
RDWA (rounds) or RDWA+Krylov procedure (triangles). All the domain of convergence is covered. For each case, the last point (coloured in black) is the edge of the convergence domain, in the present calculation conditions.}
 \label{Ic}
\end{figure}

First we determine a ``non-connected'' Floquet eigenvector, in the absence of any time-dependent absorbing potential.
In such a case, the calculation converges to an arbitrary Floquet eigenstate and not to the Schr\"odinger equation solution.
Fig. \ref{Ic} shows that the required number of iterations varies roughly linearly as a function of $E_0$ (with a few exceptional points) until divergence sets in, whatever the choosen method. The Krylov algorithm gives
the convergence up to $E_0=0.5$, whereas the simple RDWA diverges beyond $E_0=0.25$.

\begin{figure}[!ht]
 \centering
\includegraphics[width=\linewidth]{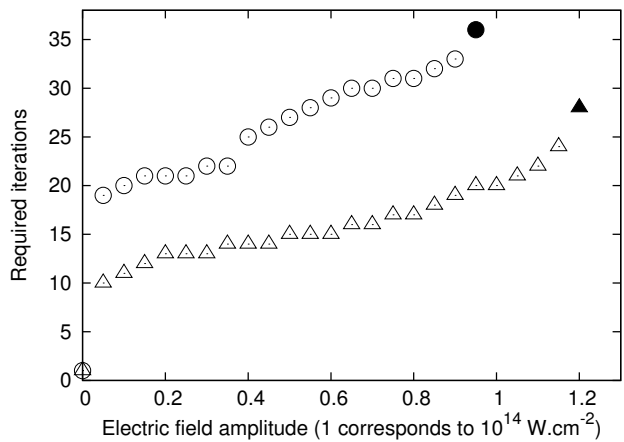}
 \caption{Number of iterations required until convergence vs electric field amplitude $E_0$, with $V_0=0.4$, using
RDWA (rounds) or RDWA+Krylov procedure (triangles). All the domain of convergence is covered. For each case, the last point (coloured in black) is the edge of the convergence domain, in the present calculation conditions.}
 \label{Ia}
\end{figure}

We then make a calculation with a quite large absorbing potential ($V_0=0.4$).
Fig. \ref{Ia} shows a relatively linear behaviour for the two methods, with a clear advantage for the Krylov method
in terms of speed but a not so marked one in terms of radius of convergence ($E_0=0.95$ for the simple RDWA, $E_0=1.20$ for the RDWA+Krylov procedure). Nevertheless we must note that the quality of the results is slightly better for the simple RDWA [$\epsilon=4.79\times 10^{-15}$ when $E_0=0.3$ as compared with $\epsilon=2.25\times 10^{-12}$, see Eq.\eqref{residue}]. Some numerical results are given in Table \ref{resultsIa} for several runs using RDWA or RDWA plus Krylov algorithm for different electric field amplitudes to check the stability of the results.

\begin{table*}[!ht]
 \centering
\begin{tabular}{lllllll}
\hline
\hline
Electric Field & Procedure & Biggest initial  & Final &&&\\ 
Amplitude $E_0$ & & residue $\epsilon$ & $P(\ket{j=0})$ &$P(\ket{j=1})$&$P(\ket{j=2})$&$P(\ket{j=5})$\\
\hline
0.3 & (A) & $4.79\times 10^{-15}$ & $0.9994647$ & $2.30363\times 10^{-4}$  & $3.8560\times 10^{-5}$ & $4.4821\times 10^{-6}$\\ 
0.3 & (B) & $2.25\times 10^{-12}$ & $0.9994673$ & $2.30355\times 10^{-4}$  & $3.8545\times 10^{-5}$ & $4.4806\times 10^{-6}$\\ 
0.5 & (A) & $2.36\times 10^{-15}$ & $0.9980043$ & $1.439085\times 10^{-3}$ & $1.86549\times 10^{-4}$ & $1.6079\times 10^{-5}$\\ 
0.5 & (B) & $1.66\times 10^{-13}$ & $0.9980010$ & $1.439042\times 10^{-3}$ & $1.86571\times 10^{-4}$ & $1.6086\times 10^{-5}$\\ 
0.7 & (A) & $5.14\times 10^{-14}$ & $0.994850$  & $4.1799\times 10^{-3}$   & $3.7229\times 10^{-4}$ & $1.5406\times 10^{-5}$\\ 
0.7 & (B) & $4.21\times 10^{-13}$ & $0.994845$  & $4.1802\times 10^{-3}$   & $3.7246\times 10^{-4}$ & $1.5413\times 10^{-5}$\\ 
\hline
\hline
\end{tabular}
\caption{Comparison of the CATM final transition probabilities $P(\ket{j},T_0)$ as a function of the electric field amplitude $E_0$ (in units of $10^{14}W.cm^{-2}$) with 2 different procedures : Simple RDWA (A), RDWA+Krylov subspace diagonalization (B).
The biggest initial residue is defined in Eq.\eqref{residue}. The absorbing potential amplitude was $V_0=0.4$.}
\label{resultsIa}
\end{table*}

\begin{figure}[!ht]
 \centering
\includegraphics[width=\linewidth]{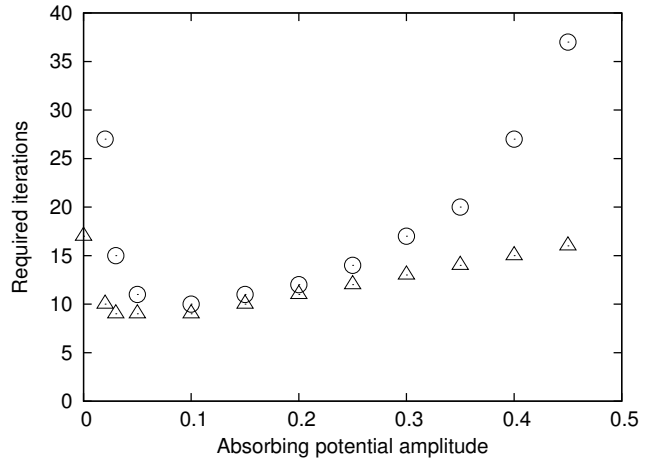}
 \caption{Number of iterations required until convergence vs absorbing potential amplitude $V_0$, with $E_0=0.5$, using RDWA (rounds) or RDWA+Krylov procedure (triangles). }
 \label{Ib}
\end{figure}

Fig. \ref{Ib} shows more clearly the influence of the absorbing operator.
We observe a divergent behaviour in the absence of the absorbing potential or for a weak amplitude ($V_0<0.05$)
when the perturbative scheme is used, while the Krylov scheme converges in the same conditions (to an eigenvector
which is not connected to the initial conditions). Then, as the absorbing operator is introduced, 
the advantage for the Krylov method disappears completely and both the methods converge in a totally
identical way. For larger amplitudes ($V_0>0.3$), the Krylov scheme again shows a significant advantage
in terms of convergence.

\begin{figure}[!ht]
 \centering
\includegraphics[width=\linewidth]{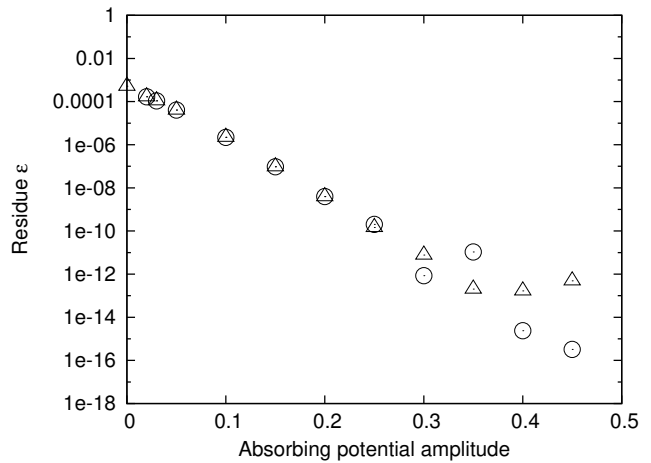}
 \caption{The $\epsilon$ of Eq. \eqref{residue} vs absorbing potential amplitude, with $E_0=0.5$, using
RDWA (rounds) or RDWA+Krylov procedure (triangles). }
 \label{Ib2}
\end{figure}

\begin{figure}[!ht]
 \centering
\includegraphics[width=\linewidth]{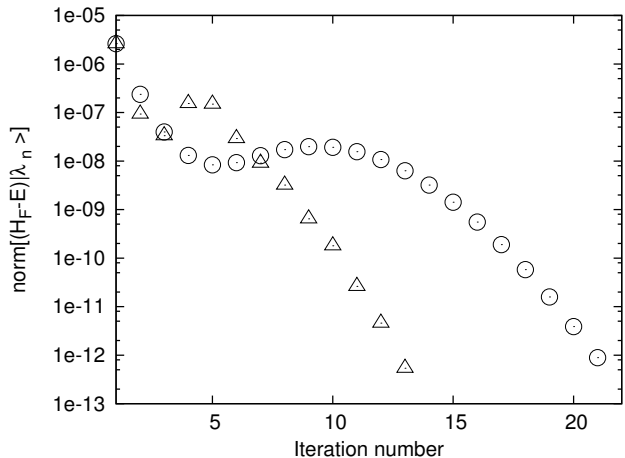}
 \caption{$\text{norm} \left[ (H_F-E_{\lambda})\ket{\lambda} \right]$ vs iteration number, 
with $E_0=0.5$ and $V_0=0.4$, using perturbative RDWA (rounds) or RDWA+Krylov procedure (triangles). }
 \label{IIa}
\end{figure}

However, we cannot study the convergence properties without keeping a check on the quality of the results.
As a measure of quality we monitor the degree of connection with the initial conditions \cite{CATM2}
by calculating the largest non-absorbed probability residue [the $\epsilon$ of Eq.\eqref{residue},
the smaller $\epsilon$, the higher the quality].
We can see in Fig. \ref{Ib2} that the residue is almost identical
for both the methods when $V_0<0.25$, but this is not a significant domain because the residue is too large to obtain accurately the solution of the Schr\"odinger equation. The residue regularly decreases as an exponential of the absorbing operator amplitude, 
but eventually the Krylov scheme shows an error which stagnates at $\epsilon\simeq 10^{-13}$ while the residue
continues to decrease for the simple RDWA. This RDWA advantage in terms of quality effectively cancels out the previous advantage for the Krylov method in terms of convergence for a strong absorbing potential.

It is also possible to compare how the two techniques approach the solution for a choosen calculation,
with $E_0=0.25$ and $V_0=0.4$. Fig. \ref{IIa} shows that the RDWA+Krylov procedure eventually approaches the solution faster than the simple RDWA procedure, after a temporary initial slowness. 

The results obtained imply that the absorbing potential best facilitates the convergence during
the search for the Floquet eigenvector when a simple RDWA perturbative method is used. 
We can explain this as follow: while the absorbing potential is necessary in order to constrain the initial condition and to obtain the solution to the Schr\"odinger equation, it also happens that under its influence the spectrum is dilated. Under these conditions a perturbative approach becomes efficient, without the need to use variational corrections through the Krylov procedure.
As a consequence the use of a Krylov subspace algorithm for the diagonalization is not always justified for the CATM, although in the absence of an absorbing potential the use of a Krylov procedure gives an advantage in calculating a non-connected Floquet eigenvector (a less important case since it does not solve the Schr\"odinger equation). When the absorbing potential is strong, the Krylov procedure converges faster but gives results of poorer quality.
Since the presence of the absorbing operator is essential in order to connect the eigenvector to the initial
condition in the CATM procedure, we conclude that the use of a Krylov subspace technique for the calculation of the eigenvector is not worthwhile.

\section{Use of an interaction representation for the hamiltonian before applying the CATM \label{reint}}

\subsection{The Gibbs phenomenom and the CATM}

From now on, all the CATM calculations are made with the RDWA procedure (cf. Appendix \ref{app1}).
Whatever the choice of the integration algorithm for the determination of the eigenvector, in the general case of the absorbing potential given by Table~\ref{derniere}, some numerical difficulties can appear, especially in the particular case of a multistep propagation (as explained in a previous article\cite{CATM3}). It is crucial that the terms $(E_j-E_l)\times \Psi^0_j /\Psi^0_l$ which are present in the column no. $l$ of the absorbing operator
should be present only during the additional time interval $[T_0,T]$ to produce the correct absorption
but must be definitely absent during the physical time interval $[0,T_0]$. For this purpose we tried to use
the Heaviside function in time, multiplying the absorbing operator by a discontinuous function
equals to one on $[T_0,T]$ and to zero on $[0,T_0]$. In practice this produced some false or diverging results;  evidently the spectral representation and the numerous FFT we used
are not compatible with the use of such discontinuous functions.
Some converged results have been presented in a previous paper \cite{CATM3} but those results could not be considered as completely satisfactory, expecially as regards the problem of the column containing $(E_j-E_l)\times \Psi^0_j /\Psi^0_l$.

To avoid this problem, two options are possible. The first one is to ensure the continuity of any time-dependent function varying too rapidly by using smooth transition functions which will soften any sudden variation by progressively turning of or turning on the sudden functions on the artificial interval $[T_0,T]$. This is conceivable but can be quite difficult to realize and may not be compatible with the aim of reproducing the exact initial condition exactly at the end of the interval without any intermediate continuity interval before the final time $T$. The second option which is much more simple is to work in the interaction representation with respect to the time-independent diagonal of the hamiltonian $H_0$ (i.e. $E_i\delta_{ii}$).

\subsection{Working in the interaction representation}

If we denote the evolution operator from $t$ to $t'$ associated with the hamiltonian $X$ by $U(t',t,X)$,
then the evolution operator associated with $H=H_0+W(t)$ can be written\cite{messiah}
\begin{equation}
 U(t,0,H)=U(t,0,H_0) \; U(t,0,V^{int})
\end{equation}
with
\begin{equation}
 V^{int}(t)=U^{-1}(t,0,H_0) \; W(t) \; U(t,0,H_0) .
\end{equation}
Practically, we apply the CATM to the transformed hamitonian
\begin{eqnarray}
\tilde{H}_{ij}(t)&=&W_{ij}(t) \times \exp \left(- \frac{i }{\hbar}(E_j-E_i) t \right) \nonumber \\
\tilde{H}_{jj}(t)&=&0, \label{eqreint}
\end{eqnarray}
with
\begin{equation}
\tilde{W}_{ij}(t)=  \mu_{ij} E(t),
\end{equation}
and after the propagation which gives the $\tilde{\Psi}(t)$ we transform this result to obtain the correct wavefunction according to
\begin{equation}
\langle j \ket{\Psi(t)}=\langle j \ket{\tilde{\Psi}(t) } \times \exp \left( -\frac{i}{\hbar} E_j t \right).
\end{equation}
Our main aim is to avoid the numerical problems previously described, because now $\tilde{H}_{ii}=0$, so that we do not need to use the problematic column in the absorbing operator. We can also use this tranformation
as a test for the CATM in order to see if our approach applies to systems with various and complicated time-dependencies everywhere in the Hamiltonian.
We would also like to know if a treatment using an interaction representation has an influence on the convergence properties of the CATM.

\begin{figure}[!ht]
 \centering
\includegraphics[width=\linewidth]{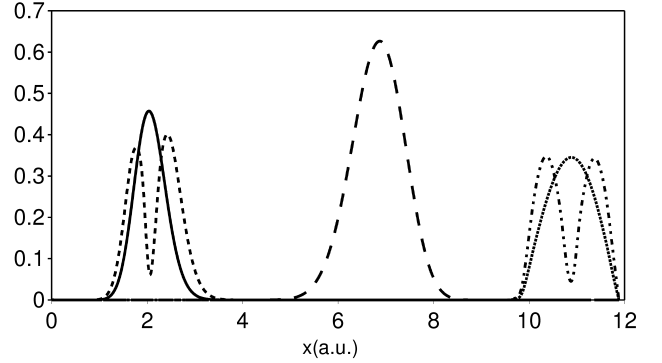}
 \caption{Spatial dependence of the modulae of the eigenstates of the field-free molecular ion $H_2^+$ : first two eigenstates of the first electronic potential curve $^2\Sigma_g^+$ on the left: $|\langle x \ket{j=0}|$ (line) and $|\langle x \ket{1}|$ (dashed line) ; first 3 pseudo-eigenstates of the second electronic potential curve $^2\Sigma_g^+$: $|\langle x \ket{102}|$ (long dashes, in the middle) $|\langle x \ket{j=100}|$ (dotted line, on the right), $|\langle x \ket{101}|$ (dotted-dashed line, on the right). These two last functions located at the edge of the grid correspond to eigenvalues with $\text{Im}(E_j)\sim -1$ and are some of those which play a very minor role in the dynamics.}
 \label{eigenstates}
\end{figure}

When we make the complete transformation of Eqs.\eqref{eqreint}, some large terms due to the non-hermiticity can appear in the hamiltonian. Thus, as we work in the eigenbasis of $H_0$, some of the basis states corresponding to the continuum have eigenvalues with quite large negative imaginary parts. This can be the source of numerical problems because of the factor $\exp \left(- \frac{i }{\hbar}(E_j-E_i) t \right)$: if $\text{Im}(E_i)\sim-1$, $\vert e^{-i(E_j-E_i)t/\hbar}\vert$ can easily reach values as big as $10^{80}$! However, these particular states are not very important in the dynamics because they are always completely localised at the edge of the grid where the radial optical potential is placed. Moreover, because of the Franck-Condon factors, the dipole moment coupling terms between such states and other states are always more than 1000 times smaller than the smallest coupling term between the states corresponding to eigenvalues with more reasonable imaginary parts. To illustrate this approximation we show the spatial form of two of these particular eigenstates in Fig. \ref{eigenstates}. We chose to neglect these problematic transitions in this case.

A second type of interaction representation can be envisaged, with respect to only the real parts of the diagonal $\text{Re}(E_i)$, with the following separation of the hamiltonian: $H=\text{Re}(H_0)+i\text{Im}(H_0)+W(t)$.
This corresponds to another transformation and we apply the CATM on the modified hamiltonian
\begin{eqnarray}
\tilde{\tilde{H}}_{ij}(t)&=&W_{ij}(t) \times \exp \left(- \frac{i }{\hbar}(\text{Re}(E_j)-\text{Re}(E_i)) t \right) \nonumber\\
\tilde{\tilde{H}}_{jj}(t)&=& i \, \text{Im}(E_j), \label{eqreintreal}
\end{eqnarray}
which gives the intermediate solution $\tilde{\tilde{\Psi}}(t)$. To obtain the solution corresponding to the original hamiltonian, we must then calculate
\begin{equation}
\langle j \ket{\Psi(t)}=\langle j \ket{\tilde{\tilde{\Psi}}(t) } \times \exp \left( -\frac{i}{\hbar} \text{Re}(E_j) t \right).
\end{equation}

\subsection{Results}

\begin{figure}[!ht]
 \centering
\includegraphics[width=\linewidth]{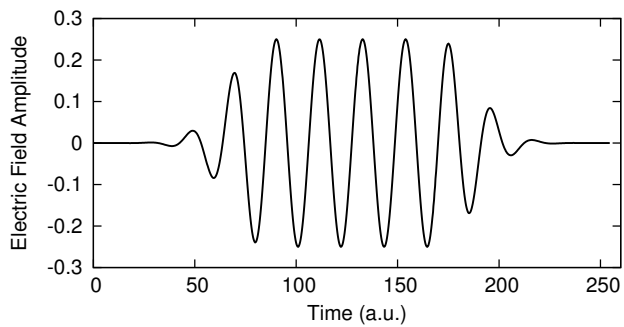}
 \caption{Laser pulse with pulsation $\omega=0.2958678$~au and total duration 254 au (i.e. 6.14 fs) with gaussian turning on and off and continuous wave during 85 au. All results in the current section relate to this pulse with varying intensity.}
 \label{champ2}
\end{figure}

In this section the pulse is slightly longer, with a gaussian turning on and of and a plateau of 85 au. It is shown in Fig. \ref{champ2}. The other parameters are exactly the same as those in section \ref{krylov}.

\begin{table}[htp]
 \centering
\begin{tabular}{llll}
\hline
\hline
Run & Inter. Repr. & Number of & Absorbing operator \\ 
 &[Eqs.\eqref{eqreint}]  & time steps & amplitude $V_0$ \\ 
\hline
(1) & no & 1 & 0.3 \\ 
(2) & yes & 1 & 0.3 \\ 
(3) & yes & 2 & 0.15 \\ 
(4) & yes & 2 & 0.3 \\ 
(5) & yes & 4 & 0.15 \\
(6) & yes & 4 & 0.3 \\
\hline
\hline
\end{tabular}
\caption{Computational parameters corresponding to the results of Table~\ref{results}. The electric field amplitude was 0.25 and the initial state is the fundamental state $\ket{i=0}$.}
\label{conditions}
\end{table}

\begin{table*}[htp]
 \centering
\begin{tabular}{llllllll}
\hline
\hline
Run & Final $P_{\text{diss}}$ & $P(\ket{0})$ & $P(\ket{1})$ & $P(\ket{2})$ & $P(\ket{3})$ & $P(\ket{9})$ & $P(\ket{16})$ \\ 
\hline
(1) & $4.6452\times 10^{-2}$ & $0.945041$ & $4.4441\times 10^{-3}$ & $2.0304\times 10^{-3}$ & $9.7800 \times 10^{-4}$& $2.600\times 10^{-5}$ & $2.268\times 10^{-6}$ \\ 
(2) & $4.6468\times 10^{-2}$ & $0.945023$ & $4.4461\times 10^{-3}$ & $2.0312\times 10^{-3}$ & $9.7825 \times 10^{-4}$ & $2.594\times 10^{-5}$ & $2.250\times 10^{-6}$ \\ 
(3) & $4.6718\times 10^{-2}$ & $0.944669$ & $4.5105\times 10^{-3}$ & $2.0503\times 10^{-3}$ & $9.8876 \times 10^{-4}$ & $2.633\times 10^{-5}$ & $2.302\times 10^{-6}$ \\ 
(4) & $4.6522\times 10^{-2}$ & $0.944973$ & $4.4453\times 10^{-3}$ & $2.0304\times 10^{-3}$ & $9.7748 \times 10^{-4}$ & $2.585\times 10^{-5}$ & $2.267\times 10^{-6}$ \\ 
(5) & $4.6497\times 10^{-2}$ & $0.944988$ & $4.4492\times 10^{-3}$ & $2.0321\times 10^{-3}$ & $9.7879 \times 10^{-4}$ & $2.596\times 10^{-5}$ & $2.257\times 10^{-6}$\\
(6) & $4.6482\times 10^{-2}$ & $0.945010$ & $4.4456\times 10^{-3}$ & $2.0310\times 10^{-3}$ & $9.7814 \times 10^{-4}$ & $2.593\times 10^{-5}$ & $2.256\times 10^{-6}$\\
\hline
\hline
 \end{tabular}
\caption{Comparison of the final transition and dissociation probabilities to bound states with the different conditions described in Table \ref{conditions}.}
\label{results}
\end{table*}

The results corresponding to computational parameters presented in Table \ref{conditions} are given in Table \ref{results}. For each run, we can change the number of steps, we can use or not use the interaction representation of Eqs.\eqref{eqreint} and we can modify the amplitude of the absorbing potential $V_0$. For the pulse shown in Fig. \ref{champ2}, the final transition probabilities to some bound states $|\langle j \ket{\Psi(T_0)}|^2$ as well as the final dissociation probability $1-\sum_{\text{bound states}}|\langle j \ket{\Psi(T_0)}|^2$ are calculated for each set of computational parameters and the numerical results are compared. Comparison between run (1) and run (2) confirms that the interaction representation is correct. Runs (3) and (4) use two steps of about 127 au (i.e. the time interval is divided into two equal parts  which are successively propagated using the CATM) and runs (5) and (6) use four time steps of about 76 au. Comparison between (2) and (3-6) indicates that the division of the calculation into several long steps using the interaction representation is valid even if we can see some minor variation in the final results. In previous articles, the influence of the absorbing potential on the results has been analysed in detail. Here we can just remark that between run (3) and run (4) the increase of the absorbing operator amplitude increases the quality of the results, in the same way as it does between runs (5) and (6).

\begin{figure}[!ht]
 \centering
\includegraphics[width=\linewidth]{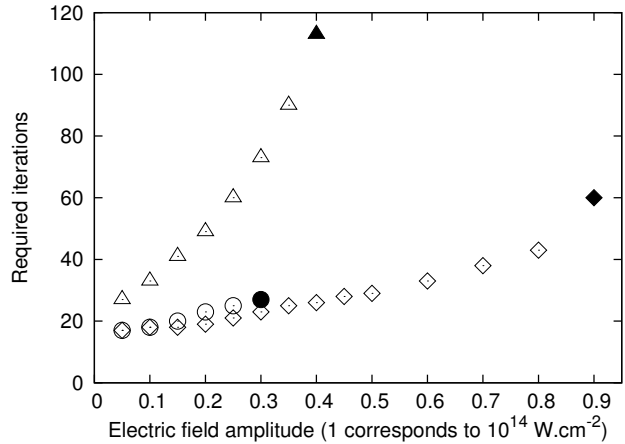}
 \caption{Number of iterations required until convergence vs electric field amplitude $E_0$, with $V_0=0.3$, using the CATM and RDWA, without interaction representation (rounds) or using an interaction representation following Eqs.\eqref{eqreint} (triangles) or Eqs.\eqref{eqreintreal} (squares). All the domain of convergence is covered. For each case, the last point (coloured in black) is the edge of the convergence domain.}
 \label{RIconvergence}
\end{figure}

Table~\ref{ABC} presents a comparison between the results given by three versions of the CATM, always working with only one global time step. It is associated with Fig. \ref{RIconvergence}, which shows the number of iterations required to obtain the convergence with the three different procedures. The first curve is the simple CATM (A), the second uses the interaction representation with respect to $H_0$ (B) [Eqs.\eqref{eqreint}] and the third corresponds to the partial interaction representation with respect to the real part of $H_0$ (C) [Eqs.\eqref{eqreintreal}].

\begin{table}[htp]
 \centering
\begin{tabular}{llll}
\hline
\hline
Electric Field & Procedure & Biggest initial  & Final $P_{diss}$ \\ 
Amplitude $E_0$ & & residue $\epsilon$ & \\
\hline
0.1 & (A) & $6.5850\times 10^{-14}$ & $9.4002\times 10^{-3}$ \\ 
0.1 & (B) & $1.2110\times 10^{-14}$ & $9.4034\times 10^{-3}$ \\ 
0.1 & (C) & $6.9398\times 10^{-13}$ & $9.4002\times 10^{-3}$ \\ 
0.2 & (A) & $1.8990\times 10^{-13}$ & $3.2795\times 10^{-2}$ \\ 
0.2 & (B) & $1.7201\times 10^{-13}$ & $3.2806\times 10^{-2}$ \\ 
0.2 & (C) & $1.3614\times 10^{-12}$ & $3.2795\times 10^{-2}$ \\ 
0.3 & (A) & $8.0980\times 10^{-13}$ & $5.9684\times 10^{-2}$ \\ 
0.3 & (B) & $8.1510\times 10^{-13}$ & $5.9704\times 10^{-2}$ \\ 
0.3 & (C) & $8.9458\times 10^{-13}$ & $5.9684\times 10^{-2}$ \\ 
0.4 & (B) & $2.1914\times 10^{-12}$ & $8.2182\times 10^{-2}$ \\ 
0.4 & (C) & $2.2416\times 10^{-12}$ & $8.2158\times 10^{-2}$ \\ 
\hline
\hline
\end{tabular}
\caption{Comparison of the CATM results as a function of the electric field amplitude (1 corresponds to $10^{14}W.cm^{-2}$) with 3 different procedures : Simple CATM (A), CATM+interaction representation [Eqs.\eqref{eqreint}] (B), CATM+interaction representation with respect to real parts [Eqs.\eqref{eqreintreal}] (C).
The biggest initial residue $\epsilon$ is defined in Eq.\eqref{residue}. The absorbing potential amplitude is $V_0=0.3$.}
\label{ABC}
\end{table}

The most simple procedure (A) shows the lowest radius of convergence, with the first difficulties appearing from an amplitude of 0.35. Using the interaction representation extends the radius of convergence and even decreases the number of iteration until we obtain a good approximation of the eigenvector. The version (B) makes easier calculations in several steps involving the absorbing operator in its most general form of Table \ref{derniere} but needs more iterations than the direct calculation (A). With the procedure (B) we note some small imprecisions due to the neglect of the large imaginary parts of the eigenvalues which are located at the edge of the grid, to avoid exponentially increasing terms. The interaction representation with respect to the real parts (C) increases the radius of convergence without needing more iterations to satisfy the convergence criteria. 
The gain in terms of convergence radius and speed is remarkable with the procedure (C). In this one-step scheme, the gain is due to the fact that the procedure replaces the integral of the couplings by Fourier transforms of these couplings, at the Bohr frequencies $(E_j-E_i)/\hbar$. We then obtain a more perturbative calculation. 


The accord between (A) and (C) is remarkable in Table~\ref{ABC}. The convergence acceleration does not automatically corresponds to a gain in term of CPU time. This is simply because this system includes couplings $\mu (x) E(t)$ with separate spatial and time dependencies, while this is no longer the case as soon as we use interaction representations. The number of FFT required increases with the use of the interaction representation, but this is specifically due to the elementary character of the studied system. It will not happen for more complicated systems, where the interaction representation will produce an appreciable gain of CPU time.

\section{conclusion \label{conclusion}}

The present article proposes a global integrator for explicitly time-dependent Hamiltonians. The addition of a time-dependent absorbing potential in order to impose the initial conditions in the Floquet treatment has the added advantage of improving the convergence in the solution of the Floquet eigenvalue equations, mainly because of the expansion of the spectrum.

The introduction of a Krylov procedure, which turns out to be very efficient for the calculation of Floquet eigenvectors without complex absorbing potential, only offers a minor advantage when the absorbing potential is present. Moreover a Krylov approach needs to store many vectors within the extended Hilbert space, which can be a problem with the CATM because the spaces in question have quite large dimensions. The absorbing potential then provides an advantage equivalent to the benefit of a Krylov subspace procedure but with a storage which is limited to one vector.

The second point of interest is that the CATM is consistent with the use of interaction representations. In this framework, the global CATM integration reduces the non-diagonal coupling amplitudes (for a relatively slow perturbation). In general, for step by step integrators, the presence of supplementary oscillations due to the interaction representation would be a supplementary difficulty, but with the global CATM scheme this facilitates the convergence.


We insist on the interest of the global approach to integrate systems which show complicated time-dependencies, possibly due to the use of an interaction representation. For half an oscillation, 4 grid points are sufficient in FFT calculation to obtain accurate results\cite{kosloff}, whereas step by step integrators can require hundreds of points, this number becoming larger if the required probabilities are small. Nevertheless, until now our test simulations on $H_2^+$ or other small systems with the CATM often take approximately the same CPU time compared to optimized step by step integrators, and depending on the conditions the CATM is sometimes faster. But our algorithm is still in work and the CPU time optimization is only one of the numerous motivation to continue the study of the CATM. Indeed this algorithm keeps its advantages already described in previous articles\cite{CATM,CATM2,CATM3}, especially the ability to the repetition of calculations, or the complete control of the accuracy of the results depending of the characteristics of the absorbing potential and of the Fourier basis set. Moreover the global structure of the CATM seems to be compatible with parallel computation, which could significantly increase the interest of this algorithm.

\begin{acknowledgments}
We acknowledge the support of the French Agence Nationale de la
Recherche (Project CoMoC).
\end{acknowledgments}

\appendix

\section{Determination of the time-dependent wave operator \label{app1}}

To obtain the results presented in this article, the constrained Floquet state has been calculated using the wave operator method. A basic RDWA iteration procedure was first applied, and then we tested the combination of this procedure with a Krylov subspace construction with approximate diagonalization. These techniques have some similarities with Davidson's method \cite{davidson}.

\subsection{RDWA iterative method}

The fundamental equation for the wave operator $\Omega$ for the Hamiltonian $H_F$ is \cite{reviewgeorges2}
\begin{equation}
 H_F\Omega=\Omega H_F \Omega .
\label{waveoperator}
\end{equation}
For the one-dimensional case, $\Omega$ is directly proportional to one eigenvector of $H_F$.
For the general case, $\Omega$ has the form $\Omega=P_0+X$ with $X=Q_0 X P_0$, $P_0$ and $Q_0$ being respectively the projectors on the active space $S_0$ and on the complementary space $S_0^{\dagger}$. Finding $\Omega$ is then equivalent to finding $X$. The effective Hamiltonian is defined by $H_{\text{eff}}=P_0H_F\Omega$.
From Eq.(\ref{waveoperator}), it is possible to get a self-consistent equation for $X$, with $H'$ an arbitrary diagonal matrix in the complementary space ($H'=Q_0H'Q_0$) and $\ket{f}$ a vector of the finite basis representation such that $Q_o \ket{f}=\ket{f}$:
\begin{eqnarray}
 &&\langle f \vert X P_0 = \left( \langle f \vert \left[ H_F-H'\right] X +\langle f\vert H_F P_0\right) \\
&& \times \left(P_0 H_{\text{eff}} P_0 - \langle f \vert H' \vert f\rangle P_0 \right) ^{-1}. \nonumber
\end{eqnarray}
One possible choice (namely the Recursive Distorted Wave Approximation) is
\begin{eqnarray}
 H'&=&\left( Q_0 (1-X) H_F (1+X) Q_0 \right) _{\text{diag}} \\
&=& \left( Q_0 (1-X) H_F Q_0 \right) _{\text{diag}}. \nonumber
\end{eqnarray}
This choice leads to the equation
\begin{eqnarray}
&& \langle f \vert X P_0 = \label{selfconsistent} \\
&& \frac{\langle f \vert H_F X P_0 - \langle f \vert (1-X)H_F \vert f \rangle \langle f \vert X P_0 + \langle f \vert H_F P_0}{
\left(H_{\text{eff}}- \langle f \vert (1-X) H_F \vert f \rangle \right) P_0}\text {, }\nonumber
\end{eqnarray}
which can be solved using several numerical procedures.

\subsection{Generalized Newton-Raphson Procedure}

We thus have to solve an algebraic equation $X=F(X)$, $X$ being a vector and $F$ the operator on the right-hand side of Eq.\eqref{selfconsistent}.
We define a trial vector $X^{(0)}$ (for example equal to the initial wavefunction delocalised over all the time interval) and then modify it iteratively by the addition of small quantities, 
\begin{equation}
 X^{(n)}=X^{(n-1)}+\Delta X^{(n)}
\label{iteration}
\end{equation}
until we obtain a sufficiently accurate satisfaction of Eq.\eqref{waveoperator}, when calculated with $\Omega^{(n)}=P_0+X^{(n)}$.
If the trial vector and the diagonal matrix $H'$ are well chosen \cite{durand,durand2}, following a classical linear procedure, the small increment is calculated as 
\begin{equation}
 \Delta X^{(n)}=F(X^{(n-1)})-X^{(n-1)}.
\end{equation}
This procedure converges almost linearly.
In the particular case of a one dimensional space $S_0$, $P_0=\vert\alpha\rangle\langle\alpha\vert$, we obtain an approximate Newton-Raphson procedure:
\begin{equation}
 \langle f \vert X^{(n+1)} \vert \alpha\rangle = \langle f\vert X^{(n)} \vert \alpha \rangle + 
\frac{\langle f \vert H^{(n)}\vert \alpha \rangle}{\langle 
\alpha \vert H^{(n)}\vert \alpha \rangle - \langle f \vert H^{(n)}\vert f \rangle}\text{,}
\label{iteration1d}
\end{equation}
with
\begin{equation}
 H^{(n)}=(1-X^{(n)})H_F(1+X^{(n)}).
\end{equation}
The result is directly tested and if the convergence criterion is satisfied we stop the calculation.


\subsection{Krylov-type procedure}

Another point of view can be adopted \cite{krylovcambridge}: the procedure of Eq.\eqref{iteration} defines a subspace of growing dimension which is spanned by the sequence of vectors $X^{(n)}$ and which contains progressively more information about the solution $X$. The growing Krylov-type basis set $\{\ket{e_i}\}$ is constructed by Gram-Schmidt orthonormalization of the sequence of the correction terms $\Delta X^{(n)}$:
\begin{eqnarray}
\ket{e_0}&=&X^{(0)} \nonumber\\
\ket{u_1}&=&\Delta X^{(1)}-\ket{e_0}\langle e_0\vert\Delta X^{(1)}\, ,\quad \ket{e_1}=\frac{\ket{u_1}}{\sqrt{\langle u_1\ket{u_1}}} \nonumber\\
&\vdots& \\
\ket{u_k}&=&\Delta X^{(k)}-\sum_{i=0}^{k-1}\ket{e_i}\langle e_i\vert\Delta X^{(k)}\, ,\quad \ket{e_k}=\frac{\ket{u_k}}{\sqrt{\langle u_k\ket{u_k}}} \nonumber
\end{eqnarray}
After a few iterations, we assume that the Krylov subspace almost contains the required eigenvector and so wish to recombine the generated approximations into something better.
After $k$ iterations, the Krylov subspace ${\mathcal{K}}^k\subset {\mathbb{C}}^{k}$ is of dimension $k$ and is spanned by the orthogonal basis $V_k\in\mathbb{C}^{n\times k}$ (containing the $\ket{e_i}$ in columns). We can then calculate a good approximation of the solution by diagonalizing the restriction of $H_F$ to the subspace ${\mathcal{K}}^k$, i.e.
\begin{eqnarray}
H_{F,k} Y &=& E Y \\
&\text{with}& \quad H_{F,k} = \overline{V}_k^T H_F V_k \, \in \,\mathbb{C}^{k\times k}  \nonumber\\
&\text{and}& \quad Y,E\,\text{(diagonal)}\, \in\,\mathbb{C}^{k\times k} \nonumber.
\end{eqnarray}
The $T$ superscript and the bar denotes the transpose operation and the complex conjugation.
On returning to the original Hilbert space, we then obtain the following approximate expression for $k$ possible eigenvectors 
associated to the eigenvalues $E$ ($N$ denotes the dimension of the extended Hilbert space):
\begin{eqnarray}
 Z_{k}=V_k Y \,\in\,\mathbb{C}^{N\times k}.
\end{eqnarray}
Among these $k$ vectors, we just have to identify the ``good'' one $X^{(k)}$, which gives the maximum value of the overlap integral of its projection at $t=0$ with the initial given wavefunction. This result is in general expected to be closer to the exact solution $X$ than the direct iterative result.

%


%


\end{document}